\documentclass[letter,twocolumn,showpacs,preprintnumbers,amsmath,amssymb]{revtex4}

\usepackage{graphicx}
\usepackage{dcolumn}
\usepackage{bm}

\begin{document}


\title{Uncertainties in Atmospheric Neutrino Fluxes}

\author{G.D.~Barr}
\email{giles.barr@physics.ox.ac.uk}
\affiliation{Department of Physics, University of Oxford,\\ 
Denys Wilkinson Building, Keble Road, Oxford, UK, OX1 3RH}

\author{T.K.~Gaisser}
\affiliation{Bartol Research Institute and Department of Physics and Astronomy,\\
University of Delaware, Newark, Delaware, USA 19716}

\author{S.~Robbins}
\altaffiliation[Now at ]{Bergische Universit\"at Wuppertal, Department of Physics, D-42097 Wuppertal, Germany}
\affiliation{Department of Physics, University of Oxford,\\ 
Denys Wilkinson Building, Keble Road, Oxford, UK, OX1 3RH}

\author{T.~Stanev}
\affiliation{Bartol Research Institute and Department of Physics and Astronomy,\\
University of Delaware, Newark, Delaware, USA 19716}

\date{21 June 2006}

\begin{abstract}
An evaluation of the principal uncertainties in the computation of
neutrino fluxes produced in cosmic ray showers in the atmosphere is
presented.  The neutrino flux predictions are needed for comparison
with experiment to perform neutrino oscillation studies.  The paper
concentrates on the main limitations which are due to hadron
production uncertainties.  It also treats primary cosmic ray flux
uncertainties, which are at a lower level.  The absolute neutrino
fluxes are found to have errors of around 15\% in the neutrino energy
region important for contained events underground.  Large
cancellations of these errors occur when ratios of fluxes are
considered, in particular, the $\nu_\mu/\overline{\nu}_\mu$ ratio
below $E_\nu=1$\,GeV, the
$(\nu_\mu+\overline{\nu}_\mu)/(\nu_e+\overline{\nu}_e)$ ratio below
$E_\nu=10$\,GeV and the up/down ratios above $E_\nu=1$\,GeV are at the
1\% level.  A detailed breakdown of the origin of these errors and
cancellations is presented.
\end{abstract}

\maketitle

\newcommand{\pt}{\ensuremath{p_T}}
\newcommand{\xlab}{\ensuremath{x_\mathrm{lab}}}
\newcommand{\eprimary}{\ensuremath{E_\mathrm{p}}}
\newcommand{\eparent}{\ensuremath{E_\mathrm{i}}}
\newcommand{\esecondary}{\ensuremath{E_\mathrm{s}}}
\newcommand{\emeson}{\ensuremath{E_\mathrm{s}}}
\newcommand{\numu}{\ensuremath{\nu_\mu}}
\newcommand{\numubar}{\ensuremath{\overline{\nu}_\mu}}
\newcommand{\nue}{\ensuremath{\nu_e}}
\newcommand{\nuebar}{\ensuremath{\overline{\nu}_e}}
\newcommand{\kl}{\ensuremath{\mathrm{K^\circ_L}}}
\newcommand{\K}{\ensuremath{\mathrm{K}}}

\section{Introduction}

The Super-Kamiokande collaboration have published detailed
analyses~\cite{sk0501064, skle} of neutrino oscillation effects based
on over 15,000 observed events induced by atmospheric neutrinos.
Neutrino oscillations have also been observed in other measurements
with atmospheric neutrinos~\cite{kamioka, soudan2, macro, minos},
accelerator neutrinos~\cite{k2k}, solar neutrinos~\cite{sno, sksolar}
and reactor neutrinos~\cite{kamland}. A crucial part of the study of
oscillation effects with atmospheric neutrinos is a detailed knowledge
of the atmospheric neutrino beam at production, before oscillations
occur.  Increasingly sophisticated calculations~\cite{calcagls,
calchkkm1d, calcfluka, calcwaltham, calcfluka2002, calcwentz,
calcfavier, calclieu, calcHKKM3d, calcbglrs} have appeared recently.
The uncertainties on the calculated flux become a limiting factor when
one uses the atmospheric neutrino beam to search for sub-leading
effects such as $\theta_{13}$ mixing, sub-maximal mixing in the
`atmospheric' sector or effects of solar
mixing~\cite{sk0604011,subdosc}.  This paper reports on a study to
enumerate the uncertainties in the neutrino fluxes due to the major
sources of uncertainty in the input to these calculations.

The main features of the atmospheric neutrino fluxes can be understood
from a discussion of their production mechanism as follows.  Cosmic
rays (about 80\% of nucleons are free protons, the rest arrive bound
in nuclei) collide with air molecules high in the atmosphere,
generating mesons which subsequently decay.  The main production of
neutrinos occurs in the decay of charged pions $\pi^+ \rightarrow
\mu^+\nu$ and the subsequent decay of the muon $\mu^+ \rightarrow
e^+\overline{\nu}_\mu\nu_e$ (and similar for antiparticles starting
with $\pi^-$).  Decay schemes involving kaons also contribute to the
higher energy neutrino fluxes.  From the main production mechanism, it
is easy to see that at low energy where muons decay before hitting the
earth, there should be roughly two muon type neutrinos for every
electron type neutrino (this is quite a good rule of thumb, because
the neutrino from the pion decay is similar in energy to the other
neutrinos due to the heavy muon).

In practice, neutrino fluxes are computed using Monte-Carlo simulation
in which the development of the cosmic ray cascade is followed
step-by-step for each track to include details of bending in the
Earth's magnetic field, of the density profile of the atmosphere and
of particle energy loss.  The influence of the Earth's magnetic field
on the primary cosmic ray flux is included using a particle
back-tracking technique to evaluate the cutoffs (see
e.g.~\cite{calcbglrs}).  This affects mostly primaries up to about
20~GeV.  The primary cosmic ray flux in the same energy range is also
modulated by the solar wind which varies with the 11-year solar cycle.

The Earth's magnetic field causes the main dependence of the fluxes on
the location on the Earth, and also produces zenith and azimuth angle
variation at each position.  The zenith angle distribution is also
affected by two other effects. (a) Higher energy muons hit the Earth's
surface and stop before decay.  This happens for vertical muons with
energy above $\sim3$~GeV which have a path length of about 20~km.
Path lengths up to 500~km are possible for horizontal muons and so
higher energy neutrinos from muon decay are present in the horizontal
fluxes.  (b) The competition between meson decay and interaction
occurs between 100 and 1000 GeV meson energy, with higher energy
particles preferring interaction due to time dilation making decay
less likely.  Since more horizontal showers develop higher up, where
the density is less, this crossover happens at higher energies than
for vertical cosmic rays.

A complication~\cite{calcfluka,lipari3d} which is dealt with in modern
Monte-Carlo calculations is the lateral spreading of the cosmic ray
showers both from transverse momentum acquired in the interactions and
decay of mesons and from bending of muons in the Earth's magnetic
field.  This makes the computation awkward as it requires generation
of particles in all directions at all points on the Earth.  The
geomagnetic field is not symmetric enough to be useful for simplifying
the problem.  A speed up trick which can be used with care is to
extend the size of the detector to cover an area extending
$\sim500$~km around the detector~\cite{calcbglrs}.
 A much larger detector increases significantly the number of 
 useful air showers.

Early calculations~\cite{calcagls,calchkkm1d} used a ``1-dimensional''
(1D) approximation in which the direction of decay and interaction
products are adjusted to lie along the trajectory of the primary
particle at its point of first collision.  In this approximation,
bending of secondaries in the geomagnetic field is neglected.  This
allows the calculation to consider only trajectories that point
directly at the detector, considerably reducing the computation time
(by about a factor of 100 in the calculation in
ref.~\cite{calcbglrs}).  One effect of making the 1D approximation is
that a geometric enhancement of sub-GeV neutrino fluxes near the
horizon~\cite{calcfluka,lipari3d} is neglected. Charge-sign dependence
of muon bending in the geomagnetic field produces differences between
neutrinos and antineutrinos up to $\sim 100$~GeV~\cite{lipari3d} which
are absent in the 1D approximation.  Both effects are difficult to see
with current detectors; the first because of the poor correlation at
low energy between the direction of the neutrino and that of the
charged lepton it produces, and the second because of the difficulty
of measuring the charge of the neutrino-induced lepton.  The geometric
effect has been found to affect only neutrinos which contain little
information about oscillations due to poor resolution of either
neutrino direction (needed to reconstruct the path length) or energy;
precisely the neutrinos which are excluded from the
analysis~\cite{skle}, see e.g.~\cite{gbarrneutrino2004}.

The effects of uncertainties on absolute fluxes is fairly
straightforward, a 10\% change in hadron production or primary flux
across the relevant regions of parameter space results in a 10\%
change in neutrino fluxes. 
 The effects of uncertainties on various ratios of fluxes, such as
 up/down, or  $\nu_\mu/\nu_e$, is much less intuitive and also of
 considerable interest.
The ratios are in principle much more
stable against the uncertainties because any change affects the
numerator and denominator in similar ways (e.g. in the $\nu_\mu/\nu_e$
ratio, both the $\nu_\mu$ and $\nu_e$ flavoured neutrinos are mainly
produced in association with muons, therefore an increase in e.g.\
pion production will increase the muon flux by a similar amount which
will increase both $\nu_\mu$ and $\nu_e$ fluxes by similar amounts).
This cancellation is absolutely vital in extracting neutrino
oscillation information from atmospheric neutrinos.  Data analyses are
constructed to take advantage of cancellation of uncertainties in the
ratios.

The main challenge in estimating the uncertainties in the computed
unoscillated neutrino fluxes is to assign errors to the different
measurements which are taken as input to the calculation.  The
dependence of the fluxes on the atmospheric density as a function of
altitude, the details of muon energy loss in the atmosphere and the
tracking in the Earth's magnetic field are found to be
small~\cite{robbins2004}.  The dominant sources of error are from
uncertainties in hadron production and following this, uncertainties
in the primary flux.  We restrict ourselves to estimating errors
from these two sources in this paper.

The hadron production uncertainty is due to the large regions of
parameter space (incident parent total energy $\eparent$, secondary
total energy $\esecondary$ (or equivalently, $\xlab$, defined as
$\esecondary/\eparent$), transverse momentum $\pt$, target atomic
weight $A$, projectile, secondary particle type) which are only
sparsely populated by measurements from accelerators.  Since
measurements of production of neutrons and $\pi^0$ are almost entirely
absent, total energy conservation is not a strong constraint.  

This sparse population of hadron production phase space also makes it
difficult to assign uncertainties on the value which has been used.
We proceed with a pragmatic approach which is described in detail in
section~\ref{sec:huncert}, to summarise, we select a given number of
regions into which to divide the phase space and then assign
independent errors to each based on the amount of existing accelerator
data in that region.  In quite a number of cases, this requires
assigning an error to the procedure of extrapolation in $\pt$,
$\xlab$, $\eparent$ or target nucleus.  This assignment has been done
by us and is somewhat subjective.  In all cases, it has been
the intent to assign errors on the basis of the level of agreement
between experimental measurements in a given region, or the amount of
extrapolation into regions where measurements do not exist.  The use
of comparison between different models to assign errors has been
avoided.  The approach described in section~\ref{sec:hdiv} has been
used to address the correlation which exists between any
mismeasurement in one region of phase space with the other regions.
To establish that this method is reasonable, several variations have
been tried and are presented in section~\ref{sec:crosschecks} with a
number of other crosschecks.

Primary flux measurements are challenging because a variety of
experimental techniques are required to cover the large energy region
of interest between 1\,GeV and 10\,TeV, the steeply falling flux as a
function of energy makes calibration a critical issue and the
experimental apparatus needs to be operated in a hostile environment
on a balloon or spacecraft.  However, several high-precision
measurements of the primary flux now exist~\cite{AMS,BESS,CAPRICE}
which resolve the historical discrepancy in the earlier data.  These
errors are included in the uncertainty estimate in a similar way to
the hadron production as described in section~\ref{sec:funcert}.

Following this, section~\ref{sec:errors} discusses the uncertainties
obtained on the absolute fluxes, the type-ratios $\numu/\nue$,
$\numu/\numubar$, $\nue/\nuebar$ and the directional ratios up/down
and up/horizontal of both muon and electron type neutrinos.  Since
most underground detectors are insensitive to lepton charge and
therefore cannot distinguish neutrino and antineutrino, throughout
this paper, the symbols $\numu$ and $\nue$ are used to refer to the
sum of neutrino and antineutrino fluxes except when explicitly used
alongside a symbol for antineutrino e.g.\ as in $\numu/\numubar$ or
$\numu+\frac{1}{2}\numubar$.  The paper continues with
section~\ref{sec:contrib} that classifies the uncertainties in the
absolute fluxes and ratios according to which regions of hadron
production phase space and primary flux are most responsible for the
uncertainties.  Various cross checks are presented in
section~\ref{sec:crosschecks} and a limitation to the cancellation
which is important when combining fluxes measured at different parts
of the solar cycle into a ratio is shown in
section~\ref{sec:wind}. Finally, concluding remarks are given in
section~\ref{sec:conc}.

\section{Previous work}
\label{sec:desc}

Several previous estimates of uncertainties have been made.  Agrawal
{\it et.\ al.}~\cite{calcagls} have extracted error estimates of
spectrum weighted moments, such as
\begin{equation}
Z_{p\pi}\;=\;\int_0^1\,{1\over \sigma_{p\pi}}
{ {\rm d}\sigma_{p\pi} \over {\rm d}\xlab }
(\xlab)^\gamma
{\rm d}\xlab.
\end{equation} 
 These arise naturally from an
analytic calculation which is possible if the flux can be approximated
by a power law dependence $E^{-{(\gamma+1)}}$~\cite{TGaisserBook} which
holds at high energies.  The authors compare spectrum weighted moments
obtained from data samples and from the hadron production models used
in the calculation.  They consider secondary pions and kaons and the
ratio between pions and kaons, but do not consider how the uncertainty
in $\pi^+/\pi^-$ or $\mathrm{K}^+/\mathrm{K}^-$ affects the ratios.
Battistoni {\it et.\ al.}~\cite{calcfluka} describe in detail the
methods by which the FLUKA Monte-Carlo hadron production generator
operates.  The authors give many examples of cross checks between
FLUKA predictions and data measurements, such as
predicting~\cite{zuccon} the measurements by AMS~\cite{AMS} from
320-390\,km altitude of the backscattering of secondary particles from
cosmic rays interacting in the atmosphere.  They assign errors of
$\sim 10$-$15\%$ on absolute fluxes and $\sim 2$-$5\%$ on the flavour
ratios.

The Super-Kamiokande paper~\cite{sk0501064} describes the issues
involved in the flux uncertainties carefully and includes estimates of
the uncertainties based on comparison of the different calculations
which have been done.  The paper also uses input on uncertainties
from~\cite{calcagls, calcfluka2002, calcHKKM3d} and the authors have
made estimates of how much their unoscillated flux model is allowed to
move.  They assign errors on the ratio $\numu/\nue$ of $\sim 3$\% for
$E_\nu < 5$\/GeV, increasing to 10\% at 100\/GeV.  Errors on
$\numu/\numubar$ and $\nue/\nuebar$ are 5\% below 10\/GeV increasing
to 25\% and 10\% above 100\/GeV for $\numu/\numubar$ and
$\nue/\nuebar$ respectively.  Errors on the up/down ratio are around
1~to 2\%.  The absolute normalisation is a free parameter in the
Super-Kamiokande oscillation fits reflecting that the uncertainties on
the absolute fluxes could be large.  

Tserkovnyak {\it et.\ al.}~\cite{calcwaltham} have performed their
calculation with different hadron production generators for
comparison.

The disadvantage of any technique which compares models as the sole
method of estimating uncertainties is the possibility of disagreements
having been tuned away as a correction to any previous comparison
between the models; i.e.\ there is the possibility that all the models
are wrong in the same way.

It is possible to use the many measurements of muon fluxes as a
validation of atmospheric neutrino flux predictions, a technique which
is used by the authors discussed above and which gives a strong
indication that the errors are in the range which they quote.  It must
be noted however that muon fluxes measured at a given altitude are
much more sensitive than neutrino fluxes measured in underground
detectors to uncertainties due to energy loss and atmospheric density.
Muons observed at sea level are roughly 1\% of all produced muons,
while neutrino fluxes measure the total number of produced muons.  The
present work concentrates on propagating experimental uncertainties in
hadron production and primary fluxes forward through the calculations
of neutrino fluxes and does not use input from muon flux measurements.

\section{Hadron Production Uncertainties}
\label{sec:huncert}

The errors on the hadron production have been estimated using only
experimental measurements which were available for the hadron
interaction models used by current atmospheric neutrino
analyses~\cite{sk0501064,skle,soudan2,macro,minos}.  More extensive
measurements covering a larger fraction of the phase space have
recently become available from E910~\cite{measE910}, 
HARP~\cite{measharp1} and NA49~\cite{measna49,measna49pc}.  
Further measurements 
with carbon targets are expected soon from HARP and MIPP~\cite{measmipp}.

The assignment of uncertainties to the different parts of the
parameter space has been done on the basis of the availability of data
and the amount of extrapolation required.  This is described in detail
in~\cite{robbins2004}.  It has largely been done in a model
independent way, but with a few guiding indicators. In particular,
when the projectile energy is in the region where resonances can be
excited, the production cross section varies rapidly across parameter
space and is difficult to extrapolate.  At higher energies, the
particle production varies more smoothly (Feynman scaling).

Since only one neutrino from any given cosmic ray shower is ever
detected, it is necessary to consider only single particle, inclusive
hadron production as we track uncertainties through the calculation.
We consider the reaction $p N \rightarrow \pi^{\pm} X$, where $p$ is the
projectile, $N$ the target nucleus $\pi^{\pm}$ the particle of interest (a
pion or kaon usually) and $X$ represents the rest of the interaction
products.  In most of the discussion of this paper, the projectile is
a proton and the target is a light nucleus such as beryllium, carbon
or aluminium, which can be extrapolated to nitrogen and oxygen, which
are the targets in the atmospheric cascade.

When a shower develops, the initial cosmic ray undergoes several
interactions in which the initial energy is split among the branches
of the shower.  Tracing backwards through the shower from the
neutrino, through the ancestors to the original cosmic ray, one can
define the branch along which any particular neutrino was produced.
At some point along the branch, there is a meson ($\pi^\pm$,
$\mathrm{K}^\pm$ or $\mathrm{K}^0$) which decays.  The neutrino may be
produced directly in this decay or via a muon.  It is the distribution
of energy, position on the globe and direction of these mesons which
govern the neutrino fluxes.  This meson will be referred to as ``the
decay meson''.

To simplify, we have limited the study of hadron production
uncertainties to the interaction in which the first meson in the
branch has been produced (i.e. its parent was a baryon).  We now
discuss two aspects which have been neglected in the uncertainties due
to this simplification. (1) It is possible in the branch that the
first meson to be produced interacts and produces a subsequent string
of hadrons before we reach the decay meson.  This depends on the local
atmospheric density and because of time dilation effects, is important
only for higher energy mesons (above $\sim200$~GeV for pions and
600~GeV for kaons).  The secondaries of such interactions will be
lower in energy, a region populated by a large number of directly
produced decay mesons from lower energy cosmic rays, so the additional
uncertainty
 because of such interactions
is small.  (2) The branch may contain a chain of
interactions in which the initial cosmic ray baryon is converted into
lower energy baryons before the decay meson is produced.  This could
happen both by near-elastic scatters or in higher multiplicity
interactions.  Some of these daughter baryons are neutrons and very
few hadron production measurements concerning either the production of
neutrons or interactions with neutron projectiles are available.
Therefore all aspects of neutrons in cosmic ray showers are obtained
from isospin arguments from proton measurements.  The uncertainty
associated with this simplification (2) requires separate study.


\begin{figure}
\begin{center}
\includegraphics[width=20pc]{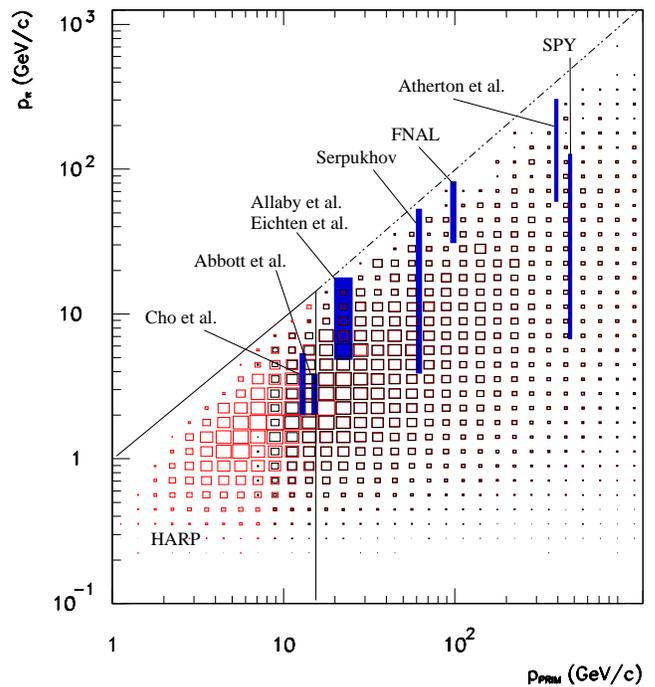}
\end{center}
 \caption{\label{fig:meassummary} (color online) Summary of
 measurements of single particle production yields as a function of
 primary energy and secondary energy.  The bands for each experiment
 represent the range of primary and secondary particle energies where
 measurements exist for at least one value of $\pt$.  The boxes on the
 plot show the contribution of the phase space to the generation of
 contained underground neutrino events as computed by the simulation.  
 The red and black boxes
 indicate the extremes of geomagnetic field effects for high and low
 geomagnetic latitude respectively.}
\end{figure}

Figure~\ref{fig:meassummary} shows a map of the ($\eparent,\emeson$)
phase space which is important for the production of neutrinos with
energies appropriate to produce contained events in underground
detectors and indicates the locations of hadron production
experiments. 

 Engel, {\it et.\ al.}~\cite{engel2000} summarize the
extent to which these measurements cover the third dimension of phase
space, transverse momentum $\pt$.  It is most important in atmospheric
neutrino interactions to understand the $\pt$-integrated yields.
Yields at specific values of $\pt$ are less important since all $\pt$
values are collected with the same probability in underground
detectors (in contrast to accelerator beams where magnetic fields are
used to focus specific regions of $\pt$ to form the beam).
Measurements over the whole $\pt$ range are important to determine the
integral.

\begin{figure}
\begin{center}
\includegraphics[width=20.3pc]{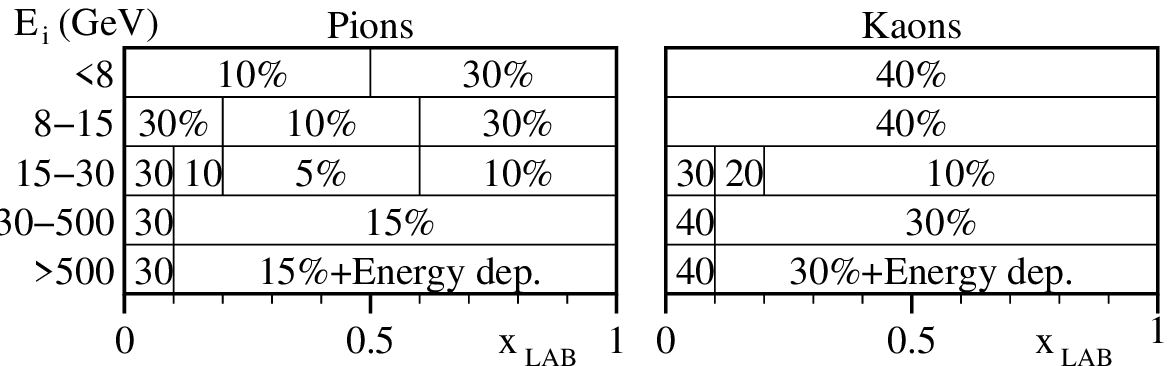}
\end{center}
 \caption{\label{fig:assigned}
  Uncertainties assigned to the production rate of charged 
  pions (left) and charged kaons (right) as a function of 
  \protect$\xlab$.  The uncertainties are shown for various 
  ranges of incident particle energy \protect$\eparent$ 
  for interactions of protons on light nuclei.}  
\end{figure}

Most of the experiments report yields for both signs of charged pions
and charged kaons.  We discuss the uncertainties on charged pion
yields first and then kaon yields.  The uncertainties assigned are
summarized in figure~\ref{fig:assigned}.  There is a series of
measurements with primary energies around
20\,GeV~\cite{eichten,allaby,abbott} with consistent measurements
spread over as much as 80\% of the $\pt$ phase space~\cite{engel2000},
allowing the hadron production to be determined without significant
extrapolation.  Above $\xlab$ of 0.6, this coverage reduces to around
50\% requiring some extrapolation.  There is only one experiment with
measurements between $\xlab$ of 0.1 and 0.2 and no measurements at all
when $\xlab$ is below 0.1.  We have therefore assigned errors of 5\%
in the region $0.2 < \xlab < 0.6$, (somewhat below the errors quoted by 
a single experiment, to account for the good agreement between measurements)
and have increased the errors to 10\% for $0.1 < \xlab < 0.2$ and to
30\% for $\xlab < 0.1$ where extrapolation in $\xlab$ is necessary.
Extrapolation to very low $\xlab$ is challenging due to the
uncertainty of how large the role of resonances is.  Above $\xlab =
0.6$, an error of 10\% is assigned due to the more limited coverage
of~\cite{eichten} in $\pt$ for $\pi^-$.  This procedure is described
in detail in ref.~\cite{robbins2004}.

Errors also include a contribution due to extrapolation between
targets.  The best solid target is carbon which like the major
components of the atmosphere contains equal numbers of neutrons and
protons.  Unfortunately, none of the
measurements~\cite{eichten,allaby,abbott} used carbon, but did measure
with Be, B$_4$C and Al.

The same procedure has been followed for $\eparent$ ranges below
8\,GeV and between~8 and~15\,GeV where measurements are available at
intermediate $\xlab$ ranges but extrapolation is required for $\xlab >
0.6$.  Extrapolation is also required for $\xlab < 0.2$ in the
$\eparent$=8-15\,GeV range.  Errors have been assigned assuming that
it is not possible to extrapolate between these different $\eparent$
ranges due to resonance effects.  The range $\eparent > 30$\,GeV is
where Feynman scaling is more effective and it is assumed that
extrapolation in $\eparent$ is possible.  The three most extensive
measurements are Barton~{\it et.\ al.}~\cite{barton} at 100\,GeV
(which includes a measurement with carbon), Atherton {\it et.\
al.}~\cite{atherton} at 400\,GeV and measurements by the SPY
collaboration~\cite{spy} at 450\,GeV.  Above this, proton-proton
measurements exist in limited regions of phase space from the ISR.  A
general 15\% uncertainty has been assigned
   to the production rates in the region $\eparent >
30$\,GeV because some extrapolation is needed between the measurements
and the $\pt$ coverage is limited from any one experiment.  Since
there are no measurements below $\xlab = 0.1$ and resonance production
in target fragmentation could be uncertain, the uncertainty is
increased to 30\% in this region.  To reflect the reliance on models
to extrapolate to high energies, where experimental data from
accelerators are unavailable, an energy dependent uncertainty $u$ is
added linearly to the 15\% (or 30\% for $\xlab < 0.1$) for
interactions in which parents are above 500\,GeV as follows.
\begin{equation}
u(\eparent) = 12.2\% \times \log_{10}\left(\frac{\eparent}{\mathrm{500\,GeV}}\right)
\label{eqn:vhighenergy}
\end{equation}
so $u$(1\,TeV)=4\%, $u$(10\,TeV)=16\% and $u$(100\,TeV)=28\%.  $u$ is
not allowed to exceed 50\% (which occurs at $\eparent=6$\,PeV).

The same procedure was used to assign uncertainties to kaons based on
charged kaon production measurements.  The uncertainties which are
assigned are shown in figure~\ref{fig:assigned}.  The kaon
uncertainties are applied to both the charged and neutral kaons in the
Monte Carlo calculation even though the level of K$^0$ and
$\overline{\mathrm{K}}^0$ production is determined through isospin
relationships in the models.  Many of the experiments which measure
pion production also measure charged kaons, however below
$\eparent=15$\,GeV there are very few kaon production measurements and
so larger uncertainties have been assigned in this region.  The same
value of $u$ as described above was added linearly to the 30\% when
$\eparent> 500$\,GeV (40\% when $\xlab < 0.1$).

Although the role of the uncertainty of secondary nucleon production
is not addressed in this study, it is worth noting that the
uncertainties are especially large in the energy region around
$\eparent=20$\,GeV.  Figure~15 of Ref.~\cite{gaisserhonda} shows a
comparison between data and various Monte Carlos; the data sets
disagree and the models do not follow the data.

The constraint imposed by overall energy conservation in the
interaction may be used to limit the size of hadron production
uncertainties provided the amount of energy carried away by neutral
particles such as neutrons and $\pi^\circ$ is estimated.  The energy
conservation constraint has not been used in the current study because
it is intended that the individual hadron production error assignments
used here can be scaled as future improvements in the measurements are
made.

\section{Uncertainty regions}
\label{sec:hdiv}

\begin{figure}
\begin{center}
\includegraphics[width=20pc]{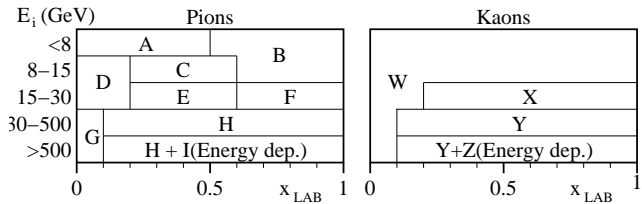}
\end{center}
 \caption{\label{fig:correl1} Uncertainty sources for hadron
 production.  The uncertainties which are applied are fully correlated
 within each region shown and completely uncorrelated between regions.
 The letters used to label each region are used on subsequent figures.
 The levels of uncertainties applied are shown in
 figure~\ref{fig:assigned}.}
\end{figure}

In order to carry out an analysis of the results
of combining uncertainties in different regions of phase space,
we want to define a set of uncorrelated sources of uncertainty
that can be varied independently.  Referring to figure~\ref{fig:correl1},
we assume that shifts within each region are fully correlated
while completely neglecting correlations between adjacent regions.
This is of course unrealistic, but we minimize the distortion
of reality by choosing the boundaries so that as far as possible
the different regions correspond to different physical effects.
Thus for pions, Regions A, D and G contain the central
region of phase space which relates primarily to multiplicity, while 
regions B, C, E, F and H represent the fragmentation region
in which the pion production is determined mainly by the valence
quarks.  Regions C and B are more closely connected to each
other, as are E and F, but we treat them separately in order
to track which regions of phase space would most benefit from
more precise measurements.

For kaons, region W represents the very poorly measured part of 
phase where resonance production is important at low energy 
(below $15$~GeV) and $s\bar{s}$ pair production in the central 
region is important at high energy.  X and Y are used to 
represent associated production (a $\Lambda K$-pair) at high energy.

The energy dependent term in both pions and kaons at very high 
energies (equation~(\ref{eqn:vhighenergy})) is varied independently,
represented by the regions I and Z.

The effect of changing the layout of these boundaries is found
to have little effect on the uncertainty determination; we defer a 
discussion of this to section~\ref{sec:crosschecks} on cross checks.

The approach chosen allows
different regions of phase space shown in figure~\ref{fig:correl1} in
the ($\eparent$,$\xlab$) plane to vary independently with the allowed
1$\sigma$ variation shown in figure~\ref{fig:assigned}.  Since both
the $\K/\pi$ ratio and the $\K^+/\K^-$ ratio are nearly as poorly
measured as the $\K$ production itself, the $\K^+$ and $\K^-$
production are varied entirely independently of each other and of the
$\pi$ production.
An uncertainty for the $\pi^+/\pi^-$ ratio of $\pm 5$\% has been
applied.  We assume that the interaction can be considered as a
combination of centrally produced {\it pairs} of pions which do not
contribute to an uncertainty in the charge ratio and a 20\%
contribution to the uncertainty from projectile fragmentation.  This
is obtained by assuming that the projectile may be excited into a
state with given isospin which subsequently decays to a nucleon and a
single pion and evaluating the different possibilities.

The number of different uncertainty sources in our analysis is 18 in
total (9 in pions, 4 in $K^+$, 4 in $K^-$ and 1 representing the
$\pi^+/\pi^-$ ratio).  It has been chosen to be roughly the same as
the number of questions and worries in the model builder's minds.  If
the number of regions is too small (e.g.\ only 2 or 3), this describes
a situation in which the shape of the hadron production is well
defined, but the overall rate is unknown; in this case, the
uncertainty cancels almost completely in the ratios of neutrino
fluxes.  If the number is too large (e.g.\ a few hundred) and all vary
independently, this describes a situation in which there are large
variations between local parts of the phase space and the
uncertainties in adjacent regions average out and again underestimates
the true uncertainty.

\section{Primary fluxes uncertainties}
\label{sec:funcert}

The primary flux uncertainties are incorporated into the uncertainties
on the neutrino fluxes in a similar way to the hadron production
uncertainties.  Gaisser, Stanev, Honda and Lipari~\cite{hamburg01}
(GSHL) have parameterized the primary fluxes as a function of energy
in solar minimum conditions, based on a compilation of a large number
of flux measurements.  Measurements up to 200~GeV/n are possible using
balloon or spacecraft mounted experiments.  At higher energies, fluxes
are determined by balloon borne emulsion-calorimeter techniques which
are less precise.  The GSHL parameterization gives the fluxes as
follows
\begin{equation}
 \Phi(\eprimary) = a\left[\eprimary + b \exp\left( c \sqrt{\eprimary} \right)\right]^{-d}
\label{eqn:stanevparam}
\end{equation}
where $\eprimary$ is the primary energy in GeV (in GeV/nucleon for
nuclear cosmic rays) and $d = \gamma+1$ is the differential spectral index. 
The parameters $a$, $b$, $c$, and $d$ are chosen
separately for the proton fluxes and for the sum of all the nuclear
fluxes.  The parameter $d$ governs the most striking feature of the
cosmic ray fluxes; the extremely steep fall-off with energy.

The uncertainties are obtained by estimating the variation required in
the parameters $a$ to $d$ to suitably cover the spread in the modern
measurements.  The values used are shown in table~\ref{tab:flux}.
\begin{table}[htb]
 \caption{\label{tab:flux}Summary of primary flux parameter variation}
\begin{center}
\begin{tabular}{lcc}
\hline
Parameter               & Proton fluxes   & Nuclear fluxes \\
\hline
$a$ (normalization)     &$ 1.49 \pm0.10$  & $  0.060 \pm0.004$   \\
$b$                     &$ 2.15 \pm0.025$ & $   1.25 \pm0.03$    \\
$c$                     &$-2.21 \pm0.02$  & $  -0.14 \pm0.02$    \\
$d$ (index) $<200$GeV/n &$ 2.74 \pm0.01$  & $   2.64 \pm0.02$    \\
\phantom{$d$ (index)} 
            $>200$GeV/n &$ 2.74 \pm0.03$  & $   2.64 \pm0.04$    \\
\hline
\end{tabular}
\end{center}
\end{table}
The uncertainty on the parameter $d$ is increased by a factor of three
for primaries above 200~GeV/n where the fluxes can not be measured
with spectrometers.

\begin{figure}
\begin{center}
\includegraphics[width=27pc]{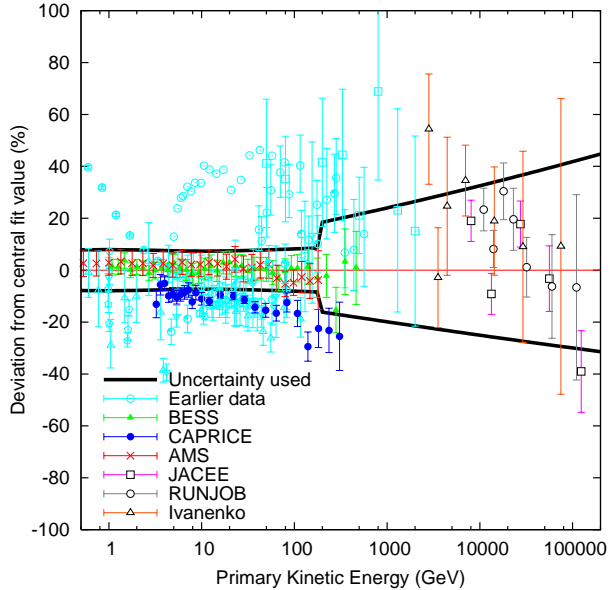}
\end{center}
 \caption{\label{fig:protonfluxes} (color online) Comparison of proton
flux measurements to the GSHL parameterization with parameters given
in Table~\ref{tab:flux} as a function of energy.  The lines show the
uncertainties on the fluxes used in this paper.}
\end{figure}

The proton flux measurements are compared on
figure~\ref{fig:protonfluxes} where the residual between the
measurements and the above parameterization are plotted against cosmic
ray energy.  Historically, there was a discrepancy among earlier measurements
of up to 50\% in the important region between 10 and 100~GeV.
Modern experiments are in much better agreement although the data of 
CAPRICE~\cite{CAPRICE} is a little lower than AMS~\cite{AMS} and 
BESS~\cite{BESS} which agree with each other very well.  Analogously 
to the hadron production, the proton flux
uncertainties are applied by using four uncertainty sources to allow
the changes to the parameters $a$, $b$, $c$, and $d$ to be applied
independently.

\begin{figure}
\begin{center}
\includegraphics[width=27pc]{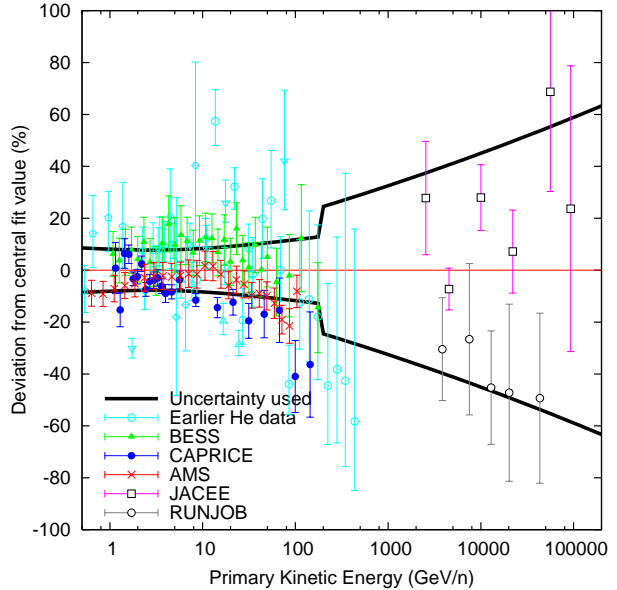}
\end{center}
 \caption{\label{fig:fluxuncert} (color online) Comparison of helium
flux measurements to the GSHL parameterization with parameters given
in Table~\ref{tab:flux} as a function of energy per nucleon.  The
lines show the uncertainties on the fluxes used in this paper.}
\end{figure}

The primary cosmic ray flux contains nucleons which are bound in
nuclei of various sizes.  For this estimate of the errors, we have
assigned errors to the fluxes to cover the spread in helium fluxes as
shown on figure~\ref{fig:fluxuncert} and used these errors as four
more uncertainty sources to represent the variation for all nuclei
other than protons.

\section{Error estimates}
\label{sec:errors}

The flux calculation proceeds by performing a Monte-Carlo calculation
with the same simulation program as in Ref.~\cite{calcbglrs} at~70
separate equally logarithmically-spaced energies between~1\,GeV
and~10\,PeV and summing up the neutrinos which are produced;
normalizing to the correct number of primary cosmic rays.  To estimate
the uncertainties, the calculation was repeated~26 times with each of
the uncertainty sources described above (18 in hadron production and 8
in primary flux) individually adjusted by 1$\sigma$ to obtain the
variation in neutrino flux as a function of neutrino energy, type and
zenith angle for each of the changes.  The hadron production
adjustment is performed by weighting the neutrinos where the first
meson is produced with the values of $\eparent$ and $\xlab$ in the
appropriate ranges.  The total uncertainty in the neutrino flux is
obtained by adding the deviations in the 26 calculated fluxes in
quadrature.  Similarly, to determine the error on a given flux ratio,
the ratio is recalculated for each of the 26 changes and these
deviations are added in quadrature.

The 1D approximation has been used to derive the error sensitivities
presented in this paper.  Although the 1D approximation affects the
values of the fluxes, it is not expected to change the sensitivity of
the fluxes to the uncertainties considered here.

\begin{figure}
\begin{center}
\includegraphics[width=20pc]{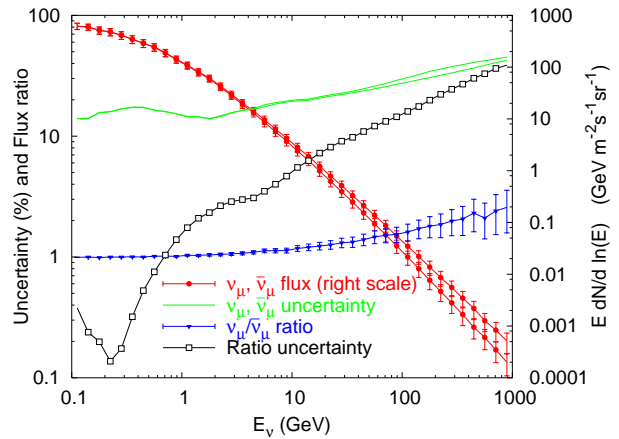}
\end{center}
\caption{\label{fig:efluxuncert} (color online) Fluxes, flux ratios
and uncertainties for muon neutrino/antineutrinos plotted as a
function of neutrino energy.  The muon neutrino and antineutrino
fluxes are shown by (red) circular points, the upper line is $\nu_\mu$
and the lower line is $\overline{\nu}_\mu$ plotted against the scale
on the right axis, the error bars represent the uncertainty from
hadron production and primary fluxes.  All of the following curves are
plotted against the scale on the left axis.  The uncertainties (in
percent) in the fluxes are shown again with the (green) lines with no
points for $\nu_\mu$ (lower line) and $\overline{\nu}_\mu$ (upper
line).  The ratio of $\nu_\mu$ flux to $\overline{\nu}_\mu$ flux is
plotted by the (blue) line with triangles and error bars indicating
the estimated uncertainty in the ratio and the uncertainties in the
ratio are shown in percent with the (black) line with square points.
The large cancellation in the uncertainties in the ratio is apparent.}
\end{figure}

Figure~\ref{fig:efluxuncert} uses the muon neutrino to antineutrino
flux ratio averaged over all zenith angles as an example to illustrate
the level to which the cancellation in ratios occurs.  The
uncertainties on the fluxes are presented in two ways, firstly the
fluxes are shown with error bars representing the uncertainties.  Note
how the neutrino flux falls steeply with energy in the same way that
the primary fluxes do.  This paper will concentrate on the
uncertainties, for a more detailed description of the features of the
fluxes themselves, see e.g.~\cite{calcbglrs}.

Figure~\ref{fig:efluxuncert} also shows the uncertainties plotted on
their own (using the left scale), which is how they will be plotted in
the later figures in this paper.  They are around 15\% at energies
corresponding to contained neutrino events which is consistent with
previous studies\cite{calcagls,calcfluka}.

In constructing the ratio of muon-neutrinos to muon-antineutrinos, the
computation takes account of the cancellation which occurs in the
uncertainties, in this case because each muon in the atmosphere is
associated with one neutrino and one antineutrino (all from the same
parent pion) and so e.g.\ any overproduction of pions will tend to
push both the numerator and denominator in the ratio upwards.  The
ratio of muon neutrino to antineutrino is shown on
figure~\ref{fig:efluxuncert} (in blue) with downward triangle markers
with error bars representing the uncertainty in the ratio.  The
uncertainty is also shown separately (as later plots in the paper will
be shown).  A large cancellation is seen at low energies and the error
on the ratio is well below 1\% around $E_\nu=200$~MeV while the errors
on the individual fluxes at this energy region is well above 10\%, a
cancellation of a factor of around 50.  At higher energies, this
cancellation becomes rapidly less powerful as some of the muons hit
the ground and so the pions only produce one neutrino.  Above 100~GeV
there is hardly any cancellation at all.

\begin{figure}
\begin{center}
\includegraphics[width=20pc]{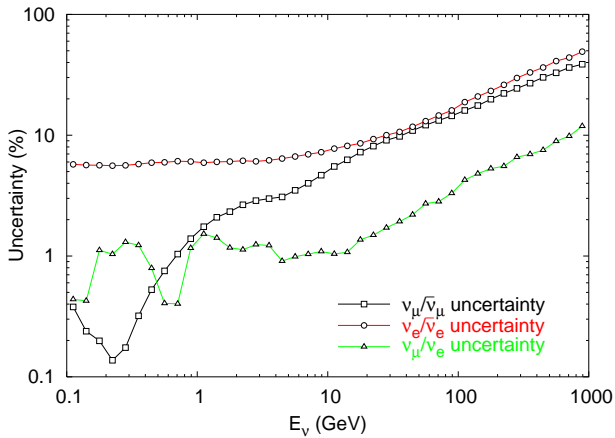}
\end{center}
 \caption{\label{fig:eflavratios} (color online) Uncertainties in
 neutrino-type ratios as a function of neutrino energy.
 $\nu_\mu/\overline{\nu}_\mu$ is shown with (black) lines with
 squares, $\nu_e/\overline{\nu}_e$ with (red) lines with circles and
 $(\nu_\mu+\overline{\nu}_\mu)/(\nu_e+\overline{\nu}_e)$ with (green)
 lines with triangles.}
\end{figure}

Figure~\ref{fig:eflavratios} presents the uncertainties on all three
neutrino type ratios averaged over all directions.  The
$\nu_\mu/\overline{\nu}_\mu$ ratio is as shown on
figure~\ref{fig:efluxuncert}.  The $\nu_e/\overline{\nu}_e$ ratio does
not show the same level of cancellation, since only a maximum of one
electron neutrino is produced from each pion.  The uncertainty on
$(\nu_\mu+\overline{\nu}_\mu)/(\nu_e+\overline{\nu}_e)$ shows similar
cancellation characteristics to $\nu_\mu/\overline{\nu}_\mu$ since a
single muon contributes to both the numerator and denominator; the
cancellation is large at low energies (where no muons hit the ground)
because muons always produce neutrinos with both flavours (in the
ratio 2:1).  At higher energies, the ratio increases, but not as fast
as $\nu_\mu/\overline{\nu}_\mu$.

\begin{figure}
\begin{center}
\includegraphics[width=20pc]{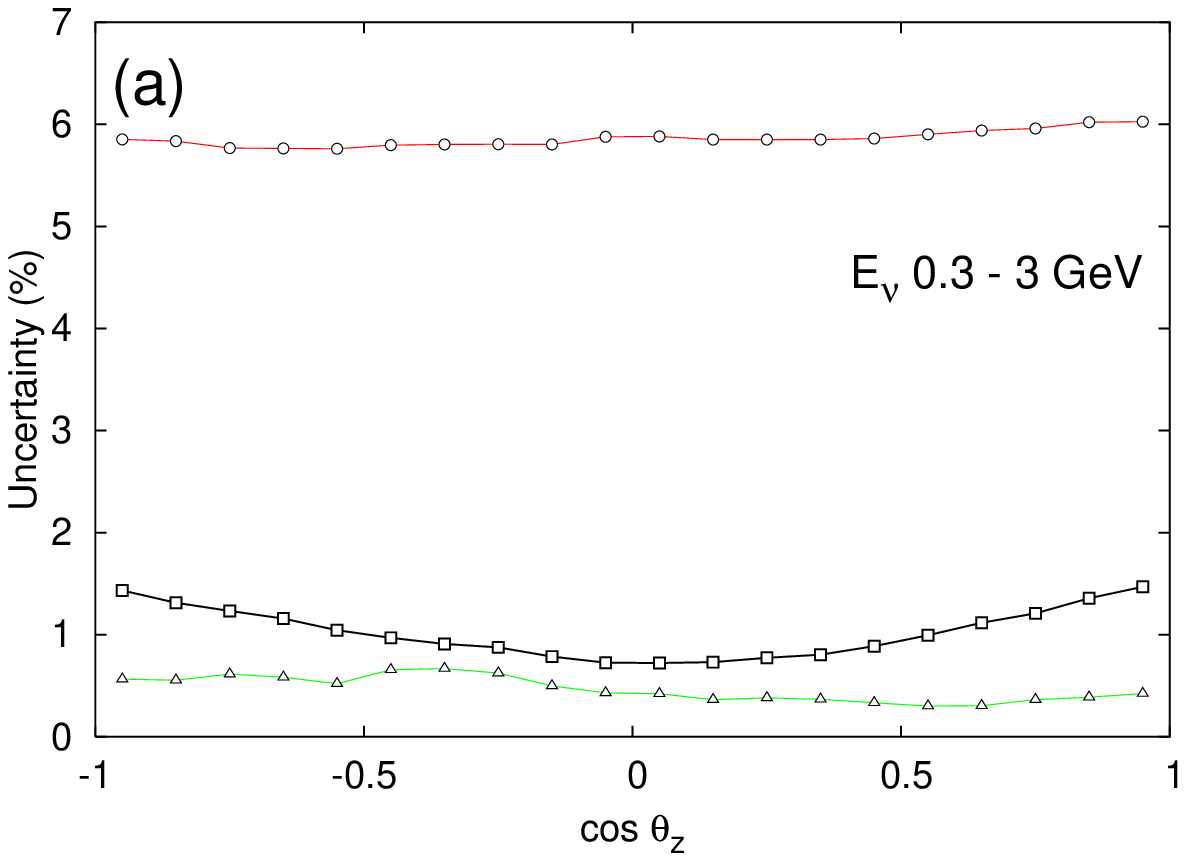}
\includegraphics[width=20pc]{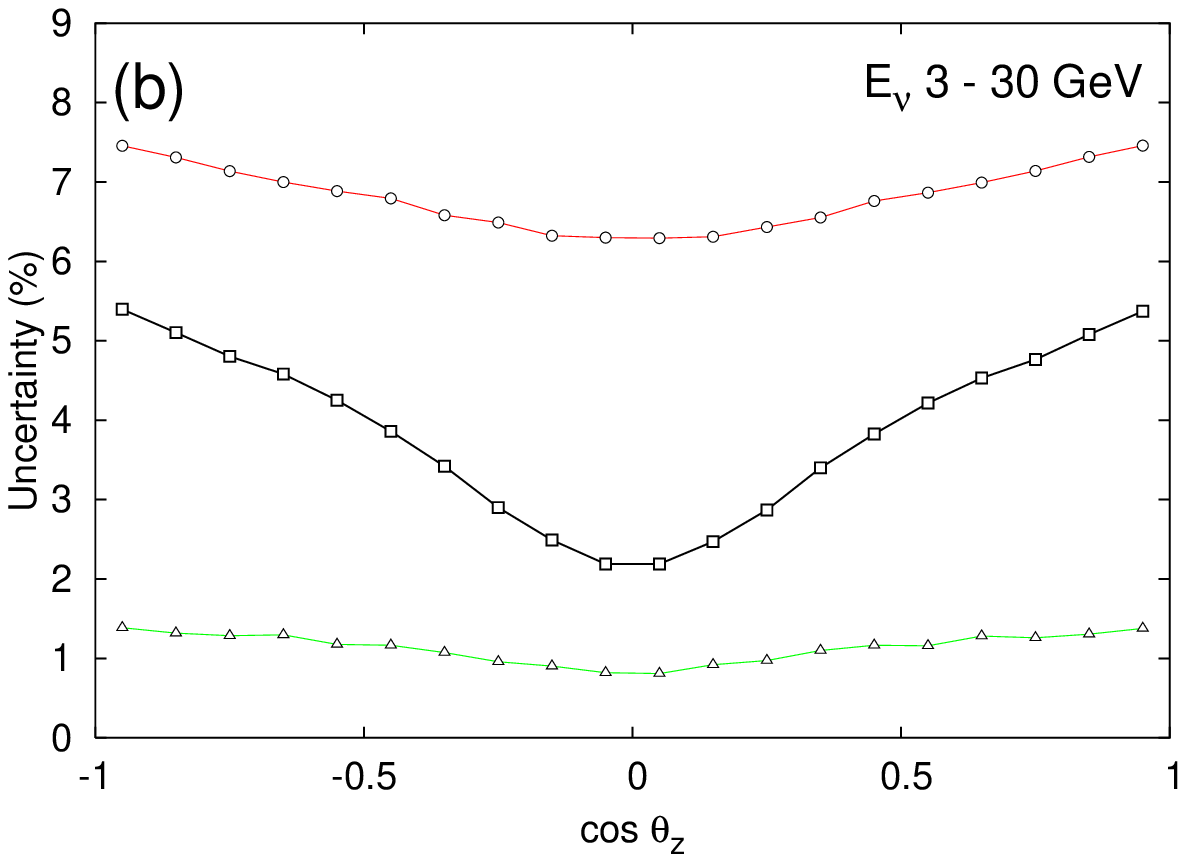}
\includegraphics[width=20pc]{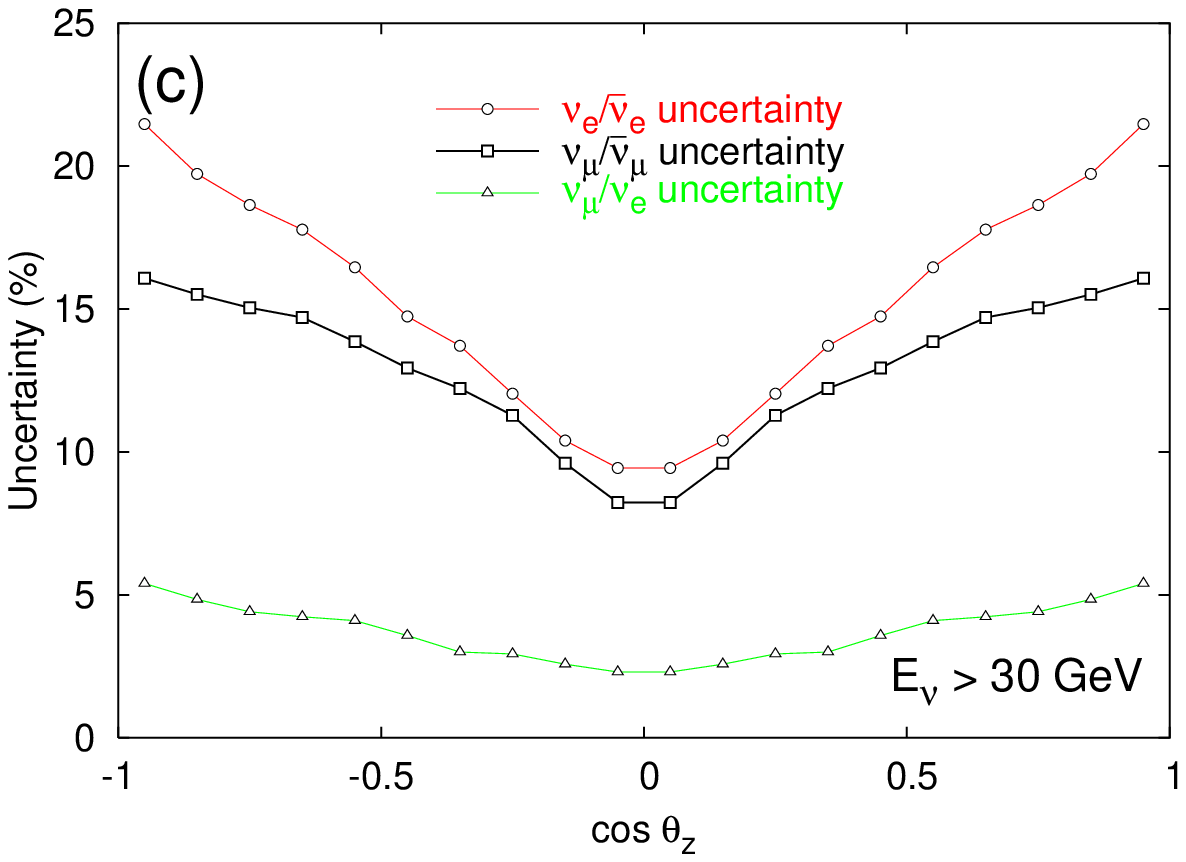}
\end{center}
 \caption{\label{fig:zflavratios} (color online) Uncertainties in
 neutrino-type ratios as a function of zenith angle for (a)
 0.3--3~GeV, (b)3--30~GeV and (c) above 30~GeV neutrino energy ranges.
 $\nu_\mu/\overline{\nu}_\mu$ is shown with (black) lines with
 squares, $\nu_e/\overline{\nu}_e$ with (red) lines with circles and
 $(\nu_\mu+\overline{\nu}_\mu)/(\nu_e+\overline{\nu}_e)$ with (green)
 lines with triangles.}
\end{figure}

Figure~\ref{fig:zflavratios} gives the variation of the neutrino type
uncertainties as a function of zenith angle for three different
neutrino energy ranges.  At low energy, below $E_\nu=3$~GeV, where the
muons do not hit the surface, the uncertainty cancellation is good in
all directions (figure~\ref{fig:zflavratios}(a)).  As shown in
figure~\ref{fig:zflavratios}(b), at intermediate energies, the
uncertainty cancellation disappears more rapidly for vertical
neutrinos than for horizontal ones as the vertical muons hit the
ground with lower energies (path length $\sim 20$~km) than horizontal
ones (path length $\sim 500$~km).  For neutrinos with energy above
30~GeV, the uncertainties on all three ratios are smaller by about a
factor of two near the horizontal than near the vertical. The
competition between interaction and decay of the mesons causes kaons
to become more important at high energy.  This occurs first near the
vertical because the local atmospheric density where the cascade
occurs is greater for vertical showers than horizontal ones.  The
large uncertainties in the ratios near the vertical reflect the
relatively large uncertainties in kaon production.

\begin{figure}
\begin{center}
\includegraphics[width=20pc]{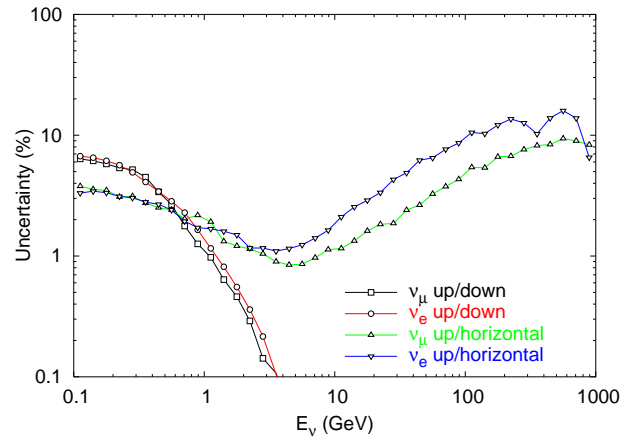}
\end{center}
 \caption{\label{fig:eUDHratios} (color online) Uncertainties in
 directional ratios as a function of neutrino energy.}
\end{figure}

Cancellation of uncertainties also occurs in ratios of neutrinos from
different directions.  These are summarized in
figure~\ref{fig:eUDHratios} which presents up/down and up/horizontal
ratios for electron type and muon type neutrinos (in each case $\nu
+\overline{\nu}$).  Since these represent neutrinos with (in
principle) identical signatures in a detector but with different
distances from the neutrino production site, these ratios are
particularly useful for neutrino oscillation studies.  For this study,
the directions are defined as follows: up is neutrinos with
$\cos\theta_z < -0.6$, horizontal is neutrinos with
$|\cos\theta_z|<0.3$ and down is neutrinos with $\cos\theta_z > 0.6$.
The calculation has used the Kamioka detector site in Japan which is
close to the geomagnetic equator (i.e. the low energy cosmic rays
locally are suppressed compared to other points on the Earth).

The uncertainty in the up/down ratio cancels exactly in the region
$E_\nu > 4$~GeV where the primary cosmic rays are above about $40~$GeV
and are too high in energy to be affected by the Earth's magnetic
field.  At these high energies, there is in principle no difference
between any two locations on the Earth and from a geometrical
argument, the upward neutrinos have the same zenith angle (but
traveling downwards) at the location of their production as at the
detector.  When $E_\nu < 4$~GeV, the geomagnetic cutoffs cause the
spectrum of primary cosmic rays to be different for the up and the
down neutrinos.  Hence, different regions of hadron production phase
space are selected and the cancellation in uncertainty is no longer
exact.  At low energies, the uncertainty in the up/down ratio is $\sim
7\%$ in agreement with earlier estimates~\cite{calcagls,sk0501064}. 
 The up/down uncertainties
are the same for both muon and electron flavoured neutrinos.

Technically, in order to avoid statistical errors from the
Monte-Carlo, it has been modified to generate neutrinos (in the 1D
mode used in this calculation) in the downward direction and then to
consider the neutrino as both downward and upwards
 by calculating the probability that the primary could
 penetrate the geomagnetic field and interact in the 
 atmosphere in the appropriate
location and direction for both directions.  This means that
Monte-Carlo statistical error on the up/down ratio comes only from the
neutrinos which pass the cutoff calculation in one direction but not
the other.  The curves shown agree well with calculations where this
modification is not implemented except for statistical fluctuations in
the uncertainty estimates when $E_\nu$ is above 4~GeV.

The up/horizontal uncertainty is also shown on
figure~\ref{fig:eUDHratios}.  Below $E_\nu = 4$~GeV, the features are
caused by the same effect as the up/down ratio shown above.  For
$E_\nu > 4$~GeV, the errors no longer cancel.  This is because of the
different atmospheric density distribution as a function of slant
depth which causes more meson reinteraction (rather than decay) for
vertical cosmic rays than for horizontal ones.  A pion which doesn't
interact has a chance to produce a high energy neutrino, whereas one
which interacts will produce a neutrino in a lower energy bin where
there are many neutrinos from lower energy cosmic rays.  Therefore the
up/horizontal ratio at high energies is uncertain because meson
reinteraction causes different regions of phase space to be
emphasized.

\begin{figure}
\begin{center}
\includegraphics[width=20pc]{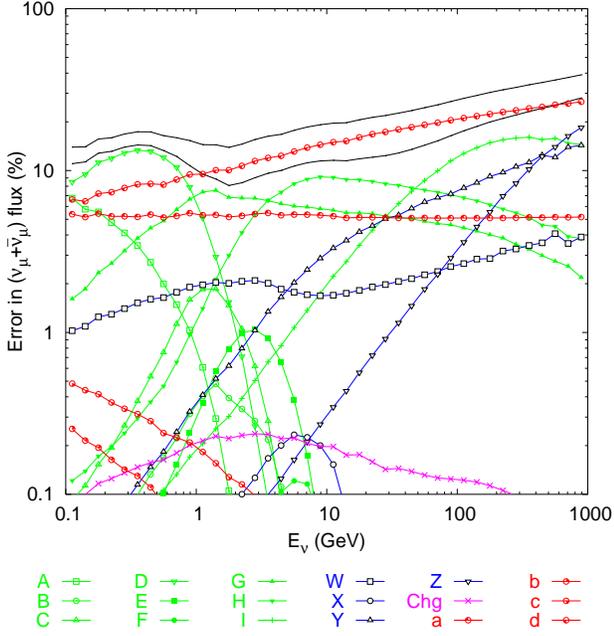}
\end{center}
 \caption{\label{fig:cmbfluxuncert} (color online) Breakdown of flux
 uncertainties (shown here for angle averaged muon neutrinos) with
 different regions of hadron production as a function of neutrino
 energy.  The capital letters in the key correspond with the hadron
 production uncertainty zones in figure~\ref{fig:correl1}
   and the one labeled `Chg' represents the pion charge ratio
   uncertainty. The
 lower case letters in the key correspond to variation of the flux
 parameters in table~(\ref{tab:flux}). See text for more information.
 The topmost thick (black) line with no points is the total error on
 the flux and the lower line of the same style is the total hadronic
 error (i.e. excluding the uncertainty from the primary flux).}
\end{figure}

\section{Contributions to the Uncertainty}
\label{sec:contrib}

\begin{figure}
\begin{center}
\includegraphics[width=19.2pc]{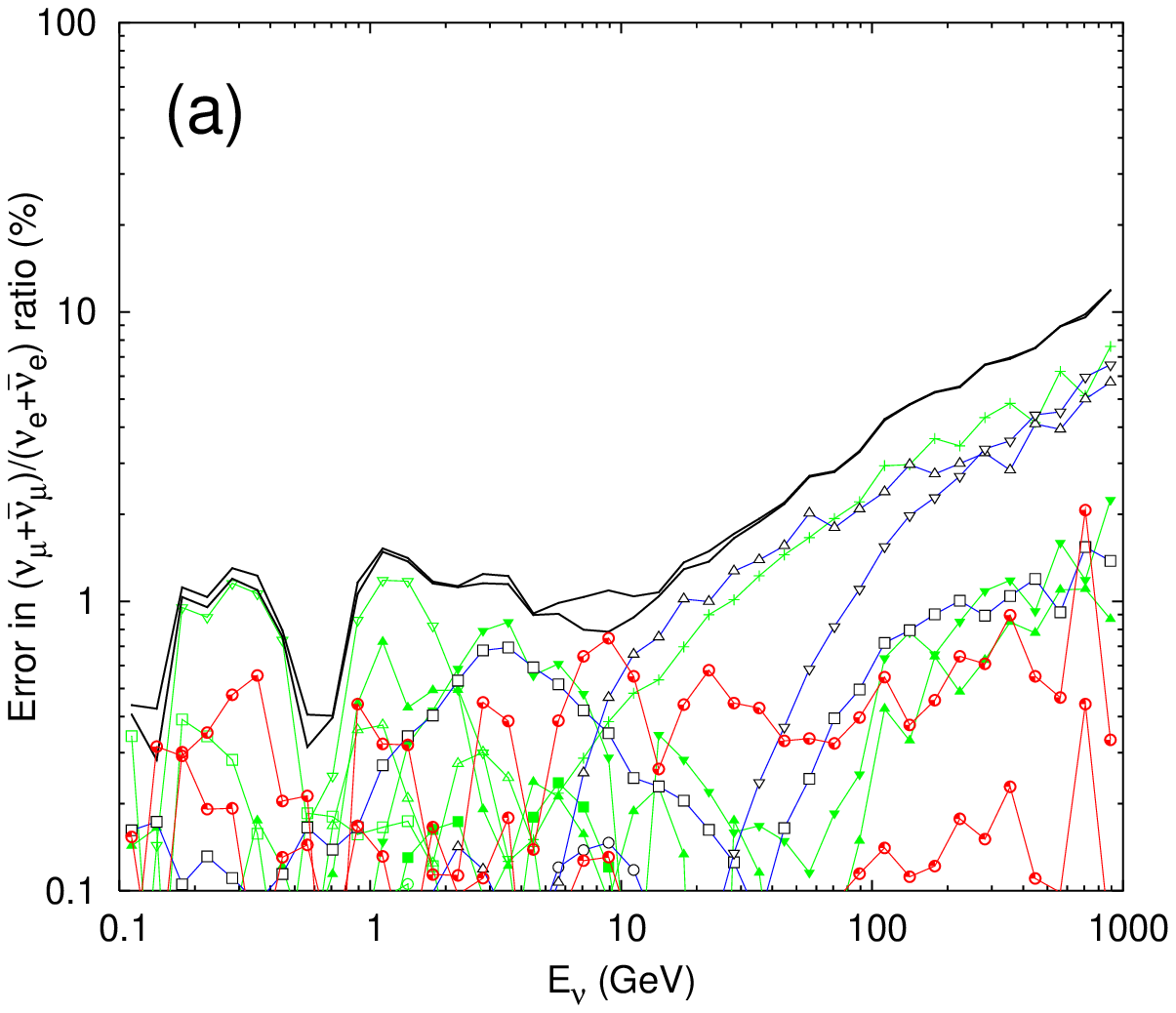}
\includegraphics[width=19.2pc]{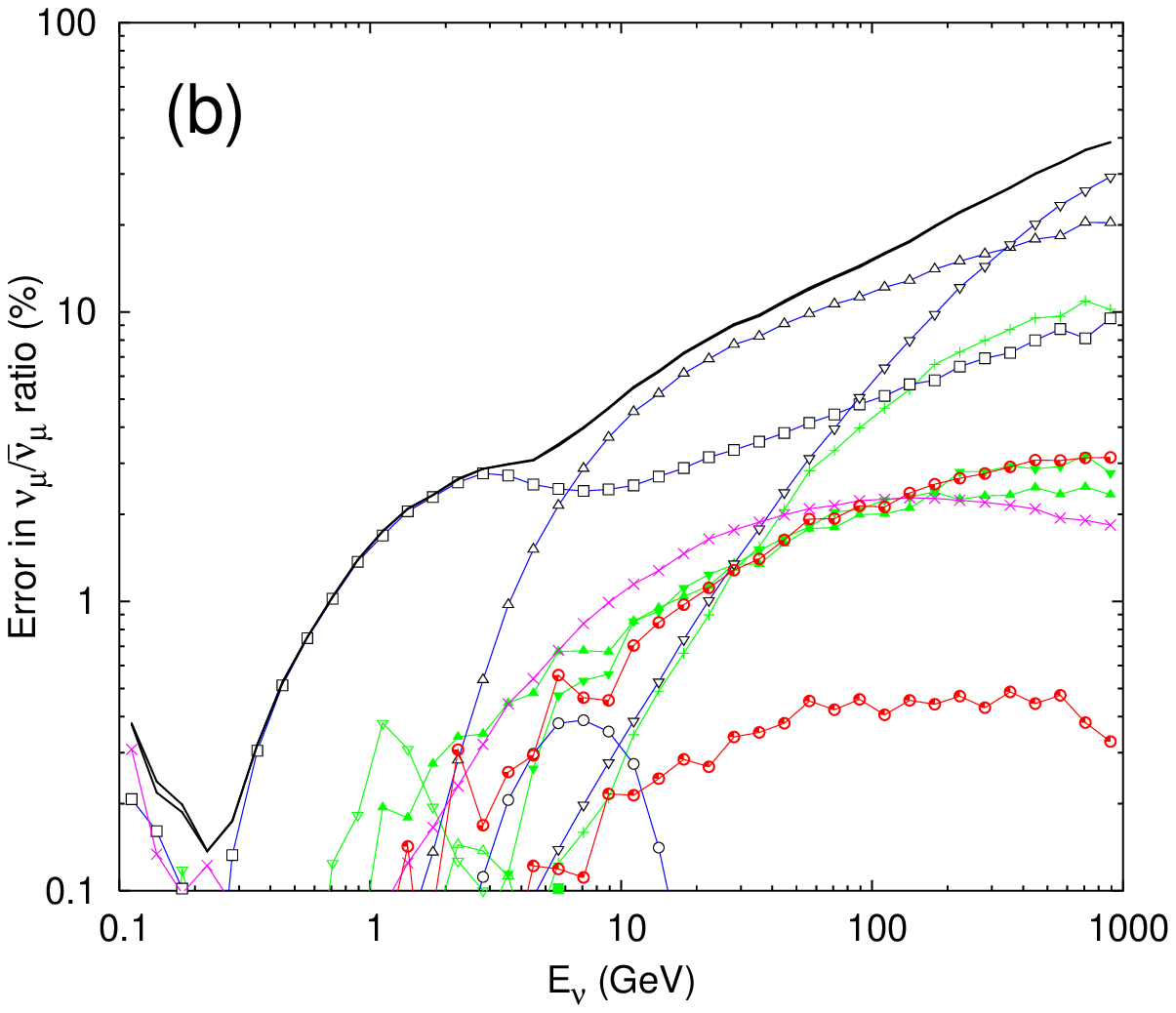}
\includegraphics[width=19.2pc]{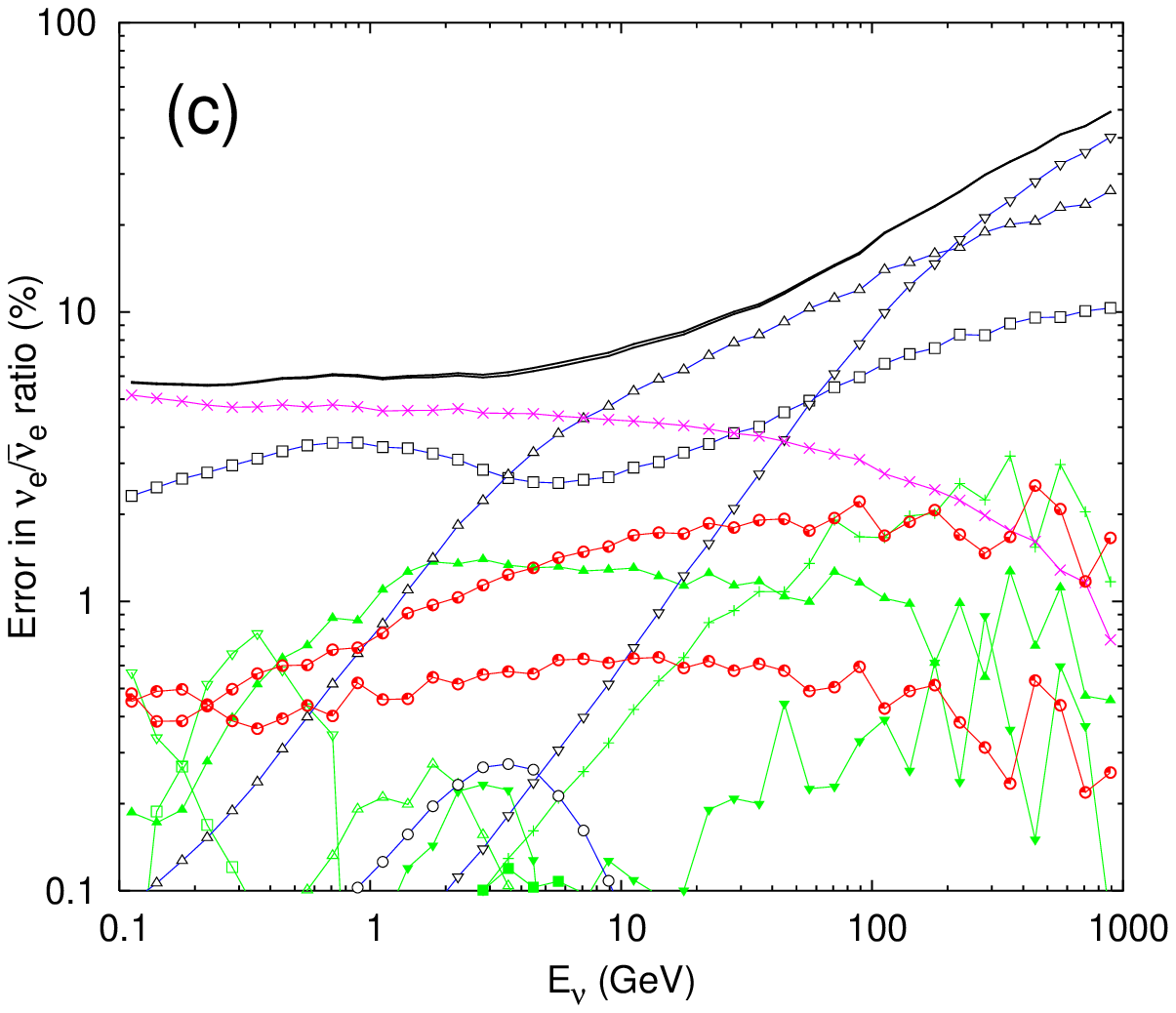}
\end{center}
 \caption{\label{fig:cmbmue} (color online) Breakdown of uncertainties
 in flavour ratios with different regions of hadron production, shown
 as a function of neutrino energy.  
 (a) $(\nu_\mu+\overline{\nu}_\mu)/(\nu_e+\overline{\nu}_e)$ ratio
 (b) $\nu_\mu/\overline{\nu}_\mu$ ratio and 
 (c) $\nu_e/\overline{\nu}_e$.
 Key as in figure~\ref{fig:cmbfluxuncert}.}
\end{figure}

\begin{figure}
\begin{center}
\includegraphics[width=20pc]{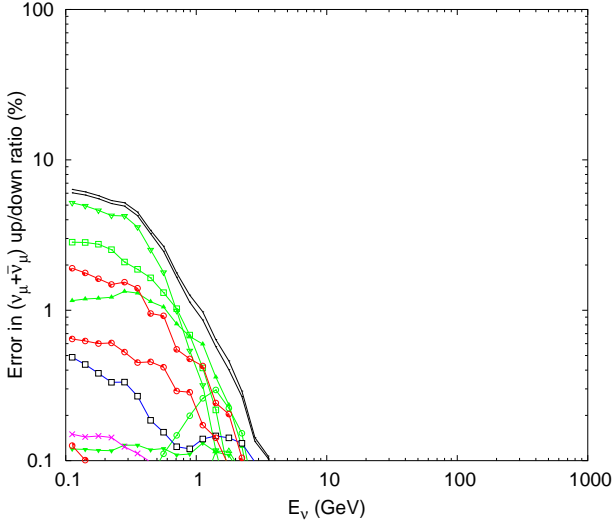}
\end{center}
 \caption{\label{fig:cmbUD} (color online) Breakdown of uncertainties
 in up/down ratio for muon neutrinos, shown as a function of neutrino
 energy, key as in figure~\ref{fig:cmbfluxuncert}.}
\end{figure}

\begin{figure}
\begin{center}
\includegraphics[width=20pc]{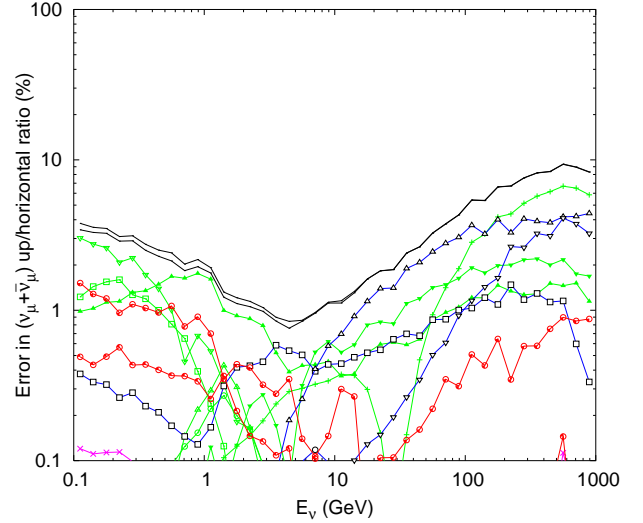}
\end{center}
 \caption{\label{fig:cmbUH} (color online) Breakdown of uncertainties
 in up/horizontal ratio, key as in figure~\ref{fig:cmbfluxuncert}.}
\end{figure}

The technique allows the contributions from the individual
uncertainties which have been inserted to be compared.  It is
interesting to identify which uncertainties have the largest effect on
any given quantity.  These are shown in
figures~\ref{fig:cmbfluxuncert} to~\ref{fig:cmbUH}.  The lines on all
these plots are the same and are described in the caption to
figure~\ref{fig:cmbfluxuncert}.  The green lines with symbols which
are either solid or with the line going through them (A -- I) are the
pion uncertainties.  The purple line with crosses (Chg) is the line
representing the pion charge ratio uncertainty (for all phase space
regions).  The blue lines with symbols which over-stamp the line (W --
Z) are the uncertainties from kaons.  Each of the four lines W -- Z
represents the combination (quadrature sum) of two independent
uncertainties for positive and negative kaons.  The phase space
regions corresponding to all these lines is summarized in
figures~\ref{fig:assigned} and~\ref{fig:correl1}.  The red lines with
quartered circle symbols (a -- d) correspond to the uncertainties on 
the primary flux parameters ($a$ -- $d$) as given in table~\ref{tab:flux}.  
Each line is the
combination of the proton and all-nuclei contribution to the
uncertainty.

Figure~\ref{fig:cmbfluxuncert} shows the breakdown of uncertainties
for the muon neutrino flux (angle averaged).  No single source of
error dominates.  Above 1~GeV the primary flux uncertainty (in
particular the value of the spectral index $d$) is important.  The
pion hadron production regions D, G, H and I are important at
respectively higher neutrino energies.  The kaon uncertainties are not
an important effect except at the very highest energies.

The three parts of figure~\ref{fig:cmbmue} show the breakdown in
uncertainties of the angle averaged neutrino type ratios.  The level
of cancellation is different for the different error contributions and
different for each of the three ratios.  In no case does the primary
flux contribution play a major role.  The
$(\nu_\mu+\overline{\nu}_\mu)/(\nu_e+\overline{\nu}_e)$ ratio shown in
figure~\ref{fig:cmbmue}(a) is mostly affected by the pion production
at low energy.  There are Monte-Carlo statistical effects in the
uncertainties below about 0.3\%.  Figure~\ref{fig:cmbmue}(b) shows the
$\nu_\mu/\overline{\nu}_\mu$ ratio which shows that the main
uncertainties are caused by kaon production.  We have perhaps been a
bit pessimistic in assigning completely separate errors to $\K^+$ and
$\K^-$ production, however this reflects the different nature of the
production of $\Lambda \mathrm{K}^+$ and $\mathrm{K}^+\mathrm{K}^-$
pairs.  The breakdown in $\nu_e/\overline{\nu}_e$ uncertainties is
shown in figure~\ref{fig:cmbmue}(c).  Since only one electron type
neutrino is produced for each pion and whether it is a neutrino or
antineutrino is almost completely defined by the sign of the pion,
this error is dominated by the pion charge ratio error assignment. We
inserted an error on $\pi^+/\pi^-$ of 5\% and the error on
$\nu_e/\overline{\nu}_e$ is essentially 5\%.  At higher energy
($E>10$~GeV), kaon production begins to dominate the error budget as
the Ke3 decay becomes the principal source of electron neutrinos.

Similarly figure~\ref{fig:cmbUD} shows the breakdown in up/down ratio
uncertainties for muon type neutrinos.  The breakdown of the errors in
up/down ratios for electron neutrinos is almost identical.  The
largest source of errors comes from hadron uncertainty source D
followed by A; i.e. the hadron production regions at low
$x_\mathrm{LAB}$.  The recent hadron production measurements cover
this region, so the precision to which this ratio can be predicted
should improve.

Figure~\ref{fig:cmbUH} shows the up/horizontal flux ratio uncertainty
breakdown for muon type neutrinos.  At low energies, the contributions
are similar to the up/down ratios.  At high energies, all the
uncertainty sources G, H, I, W, Y and Z representing the uncertainties
in hadron production with high energy parent particles are present,
indicating that the different slant depth has caused a reduction in
the cancellation of uncertainties.
  
The breakdown for electron type neutrinos in the up/horizontal ratio
is similar with somewhat higher contributions from each of the hadron
production uncertainty sources as shown on
figure~\ref{fig:eUDHratios}.

\begin{figure}
\begin{center}
\includegraphics[width=20pc]{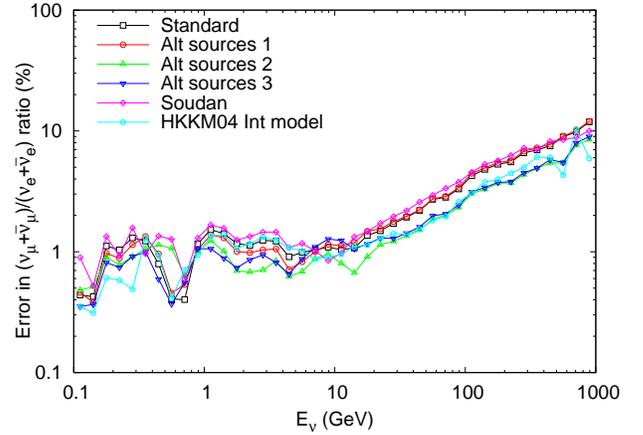}
\end{center}
 \caption{\label{fig:changes} (color online) Stability of the
 uncertainty estimates as a function of different choices of
 uncertainty sources ((1), (2), (3)), detector location (Soudan) and
 base hadron production generator.}
\end{figure}

\section{Cross Checks}
\label{sec:crosschecks}

To check the stability of the uncertainty estimates presented in this
paper, they were recalculated a number of times to quantify the
effects of modifications.  First, the layout of the uncertainty
sources within the hadron production phase space was changed in three 
different ways.
The effect on the $\nu_\mu/\nu_e$ ratio is shown in
figure~\ref{fig:changes}.  In all three cases, the size of the uncertainties
themselves were maintained the same as in the standard calculation as
shown on figure~\ref{fig:assigned}.  Alternative~1 divides up the
phase space in the same way as the standard uncertainty sources shown
on figure~\ref{fig:correl1} but to reflect that the uncertainties in
the measurements are also present in the regions where they are
extrapolated, each region of ($\eparent,\xlab$) is influenced by up to
two separate sources added in quadrature, the first is at the level of
a region where measurements are made and the second reflects the
additional error we added to account for extrapolation.  For example,
the region 8\,GeV$<\eparent<$15\,GeV and $\xlab <$0.2 is comprised of
an uncertainty source with 10\% uncertainty reflecting the measurement
error and 28\% due to extrapolation leading to a total error of 30\%.

\begin{figure}
\begin{center}
\includegraphics[width=20pc]{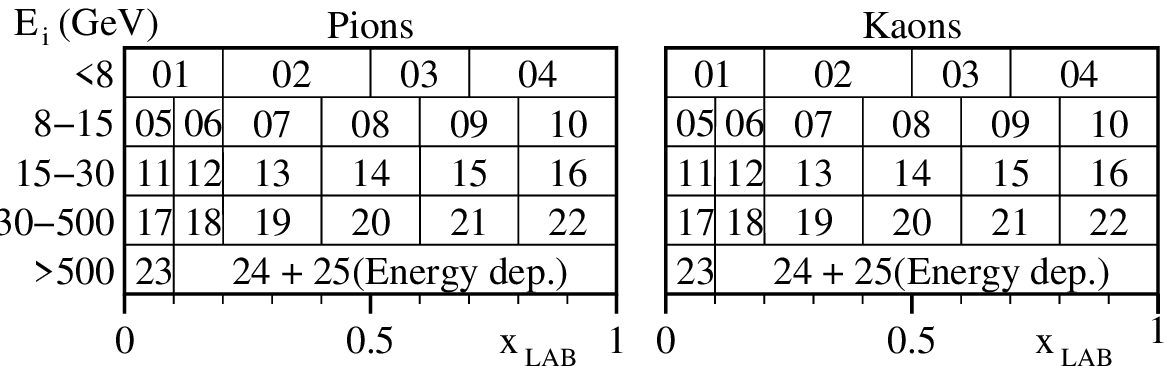}
\end{center}
 \caption{\label{fig:correl2} Uncertainty sources used in the
 alternative 2 hadron production weighting scheme.  The sources 01-25
 allow more independence to different values of $\xlab$ than the
 standard scheme and have full correlation between pions and kaons in
 the same region of ($\eparent,\xlab$).}
\end{figure}

\begin{figure}
\begin{center}
\includegraphics[width=20pc]{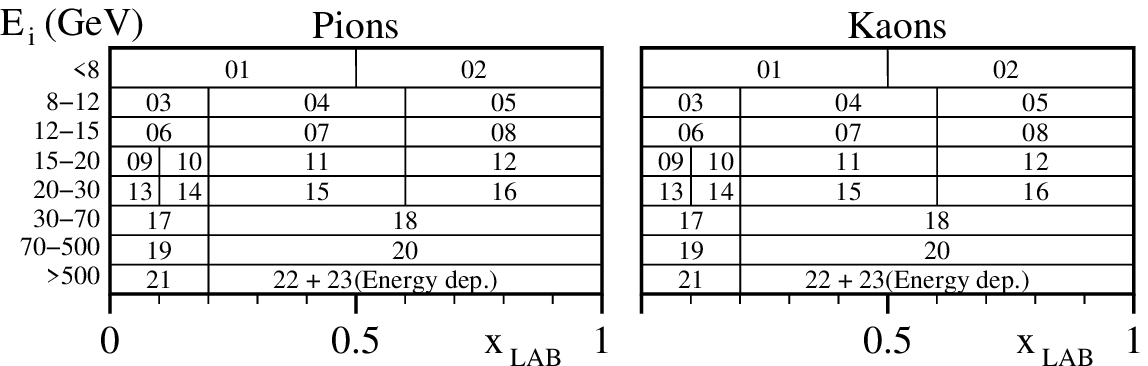}
\end{center}
 \caption{\label{fig:correl3} Uncertainty sources used in the
alternative 3 hadron production weighting scheme.  The sources 01-23
allow more independence to different values of $\eparent$ than the
 standard scheme and have full correlation between pions and kaons in
 the same region of ($\eparent,\xlab$).}
\end{figure}

Alternative~2 rearranges the uncertainty sources to be more finely
divided in $\xlab$ and uses the same uncertainty sources for pions and
kaons.  It is shown in figure~\ref{fig:correl2}.  Similarly,
alternative~3, shown in figure~\ref{fig:correl3} divides phase space
more finely in $\eparent$.  All three schemes include the uncertainty
sources for the primary fluxes and the $\pi^+/\pi^-$ ratio.
Alternative~1 handles the $\mathrm{K}^+/\mathrm{K}^-$ in a similar way
to the standard method and alternatives~2 and~3 include the
$\mathrm{K}^+/\mathrm{K}^-$ ratio uncertainty with a single source.
As seen on figure~\ref{fig:changes}, the uncertainty on the
$\nu_\mu/\nu_e$ ratio is independent of which scheme is used.  This is
also true in general for the other ratios.  

Figure~\ref{fig:changes}
also shows two other cross-checks which do not significantly affect
the error estimation.  The uncertainty analysis was run with a
different detector location, at Soudan in North America near one of
the geomagnetic poles.  

Since the analysis operates by applying
weights to different regions of phase space it could give misleading
results if, for example, part of the phase space is not used at all by
the hadron production generator.  We therefore repeated the analysis
using the hadron production generator from the HKKM04
calculation~\cite{calcHKKM3d} to verify that the same results were
obtained as shown on figure~\ref{fig:changes}.

A further cross check was to perform the combination of errors from
different hadron production regions using a statistical method.  The
standard method moves each uncertainty source by $1\sigma$, computes
the movement of each ratio and adds the effects of each uncertainty
source in quadrature.  The statistical method performs 500
calculations each with randomly assigned uncertainties to all the
uncertainty sources.  The error on the fluxes and ratios was then
determined from the widths of the distribution of the 500 different
results.  This method is equivalent and gave identical results within
the limits of Monte-Carlo statistical fluctuations.

\section{Solar wind}
\label{sec:wind}

\begin{figure}
\begin{center}
\includegraphics[width=20pc]{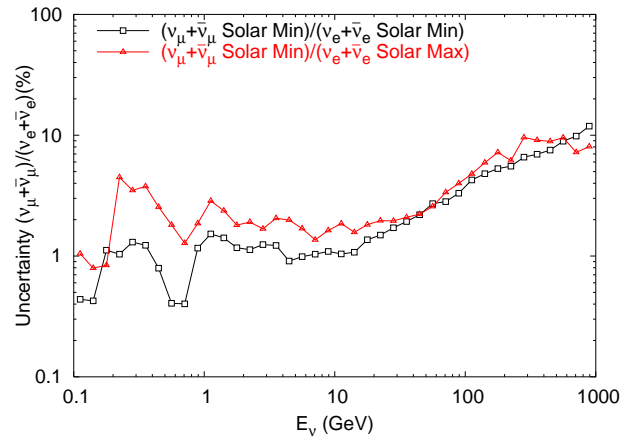}
\end{center}
 \caption{\label{fig:specials2} (color online) 
 Increase in muon neutrino to electron
 neutrino flux uncertainty when comparing fluxes from different parts
 of the solar cycle at Soudan.}
\end{figure}

As seen in the distributions above, the uncertainties in flux ratios
can be reduced significantly by cancellations.  So far, all ratios
shown assume that the ratios which are formed are taken
simultaneously, i.e. no uncertainty is added if the two components of
a ratio were taken at different times in the solar cycle where the
solar wind changes the primary spectrum below cosmic ray energies
of~$\sim 10$\,GeV.  The MINOS experiment~\cite{minos} is situated in
the Soudan iron mine in Northern Minnesota, USA at the same location
as the Soudan-2 experiment.  Ref.~\cite{minos} which studies muon
flavoured neutrinos uses $\nue$ data from ref.~\cite{soudan2} to
normalize the unoscillated fluxes from the Monte-Carlo
calculation~\cite{calcbglrs}.  Since the Soudan-2 data and the MINOS
data are taken at different parts of the solar-cycle, the large
cancellation in the $\nu_\mu/\nu_e$ ratio shown in
figure~\ref{fig:eflavratios} is not so exact.
Figure~\ref{fig:specials2} compares the uncertainty on the
$\nu_\mu/\nu_e$ ratio between the standard calculation and one in
which the $\nu_\mu$ is taken at solar minimum conditions and the
$\nu_e$ at solar maximum conditions.  The uncertainty cancellation is
not so good and is about $2.5\%$ averaged over the energy region
appropriate for the MINOS data sample.
\section{Conclusions}
\label{sec:conc}

A detailed survey of the main uncertainties involved in the
computation of the production of neutrinos in the atmosphere has been
presented.  The major contributions which come from hadron production
and the primary fluxes have been addressed.  Regions where
uncertainties exist and the amount of uncertainty in each region have
been assigned based on the data which exists and not on any agreement
or otherwise between different models (which may have been initially
tuned on the same data or compared and adjusted to enhance agreement).
An attempt to account for correlations of uncertainties across
hadron production phase space has been employed and variants tested
showing that the main conclusions of this paper do not depend on the
particular scheme used.

The uncertainties on the fluxes are around 15\% in agreement with
previous evaluations.  There is considerable cancellation of
uncertainty when taking ratios of fluxes, either from different
directions or of different neutrino types.  These are of similar size
to previous estimates~\cite{calcagls,calcfluka,sk0501064}, 
but are found to vary considerably with
neutrino energy.

The breakdown of the uncertainties reveals that there is no single
  hadronic
source of uncertainty which dominates.  The production of pions at low
energy which is currently being measured by several modern experiments
is most important in the up/down and up/horizontal ratios for
contained neutrinos.  The knowledge of kaon production is important
even for neutrinos as low as 4~GeV, particularly in the muon neutrino
to antineutrino ratio.

As described in the paper, the method of assigning and applying the
hadron production uncertainties to this problem is not unique or
complete.  One major omission with this method is that we have only
included uncertainty at one point along the path to production of the
neutrino, namely, at the production of the first mesonic ancestor of
the neutrino.  Another potential source of uncertainty is in the
baryonic part of the shower. A primary proton may undergo several soft
elastic or near elastic collisions, on it's journey to the collision
where the decay meson is produced.  Low energy proton production is
especially badly measured and models disagree
considerably~\cite{gaisserhonda}.  In addition, data on neutron
production and also on interactions with neutron parents are very
scarce.

The large degree of cancellation in the flux ratios, in particular
seen in the primary fluxes gives us encouragement that these baryonic
shower uncertainties are not going to be a major effect.  Since they
occur upstream of the hadron uncertainties we consider, they could be
represented as an increase in flux uncertainty by redefining 'flux' to
mean the spectrum of the baryons at the point of production of the
first-meson.

\begin{figure}
\begin{center}
\includegraphics[width=20pc]{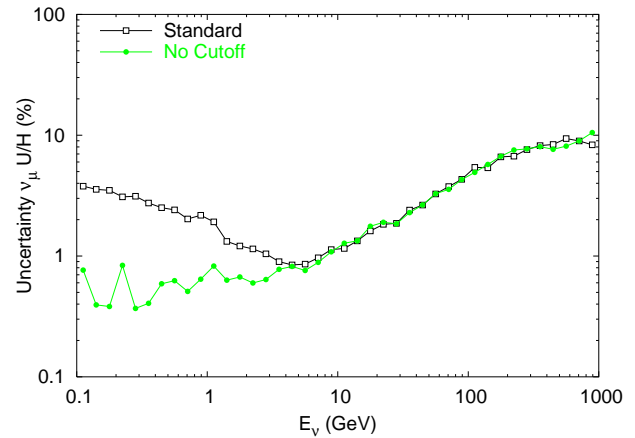}
\end{center}
 \caption{\label{fig:specials1} (color online) 
 Line with open squares shows the
 standard uncertainty in the up/horizontal $\nu_\mu$ flux ratio and
 line with closed circles the uncertainty with the effect of the
 Earth's magnetic field removed.}
\end{figure}

The cancellations in uncertainties in flux ratios are a central part
of any analysis of neutrino oscillation using cosmic rays.  This study
shows the details of this cancellation as a function of neutrino
energy and for the first time identifies the sources of the
uncertainties which remain.  There are two major effects which are
identified in the direction ratios which cause the cancellation to be
incomplete.

The first effect is observed when the numerator and denominator of the
ratio originate from cosmic rays which have different geomagnetic
effects.  This causes some regions of hadron production phase space to
be only included in either the numerator or the denominator, so the
uncertainty of that part of hadron production phase space does not
cancel.  This is apparent in the up/down ratio at low energy shown on
figure~\ref{fig:eUDHratios}.  At high energy, where the cosmic rays
are not removed by the geomagnetic field, the cancellation in the
up/down ratio is exact.

The second effect is that if the slant depth in the atmosphere is
different for the numerator and denominator (e.g. the up/horizontal
ratio, but this applies to comparing any two different general zenith
angle directions), then the balance between meson decay and
interaction is different and cancellation is incomplete.  This is
illustrated on figure~\ref{fig:specials1} which compares the usual
up/horizontal uncertainty with one where all geomagnetic field effects
have been removed.  At low energy, the uncertainty is zero since all
directions are now equivalent, but at higher energy, the uncertainty
is as large as in the full uncertainty calculation.

\subsection*{Acknowledgments}

The work of TKG and TS is supported in part by the U.S. Department of
Energy under DE-FG02 91ER 40626.  The authors would like to thank
P.~Lipari for many useful discussions concerning the uncertainties of
neutrino fluxes.

\bibliography{uncertainflux}

\end{document}